\documentclass[aps,prl,twocolumn,groupedaddress,showkeys]{revtex4}

\usepackage{graphics}

\usepackage{epsfig}

\begin{document}
\bibliographystyle{unsrt}

\title{Time-invariant person-specific frequency templates in human brain activity}

\author{Itai Doron$^{1}$, Eyal Hulata$^{1}$, Itay Baruchi$^{1}$, Vernon L. Towle$^{2}$, and Eshel Ben-Jacob$^{1}$}
\affiliation{$^{1}$School of Physics and Astronomy, Raymond \&
Beverly Sackler Faculty of Exact Sciences, Tel-Aviv University,
Tel-Aviv 69978, Israel} \email{eshel@tamar.tau.ac.il}
\affiliation{$^{2}$Department of Neurology, BH2030, The University
of Chicago, 5841 South Maryland Avenue, Chicago, Illinois 60637, USA} 


\begin{abstract}
The various human brain tasks are performed at different locations
and time scales. Yet, we discovered the existence of
time-invariant (above an essential time scale) partitioning of the
brain activity into personal state-specific frequency bands. For
that, we perform temporal and ensemble averaging of best wavelet
packet bases from multi-electrode EEG recordings. These personal
frequency-bands provide new templates for quantitative analyses of
brain function, e.g., normal vs. epileptic activity.
\end{abstract}
\pacs{} \keywords{wavelet packets, best basis, electrocorticogram}
\maketitle
\begin{figure}[btp]
\centerline{a \hspace{4cm} b \hspace{2.6cm}} \centerline{
\psfig{figure=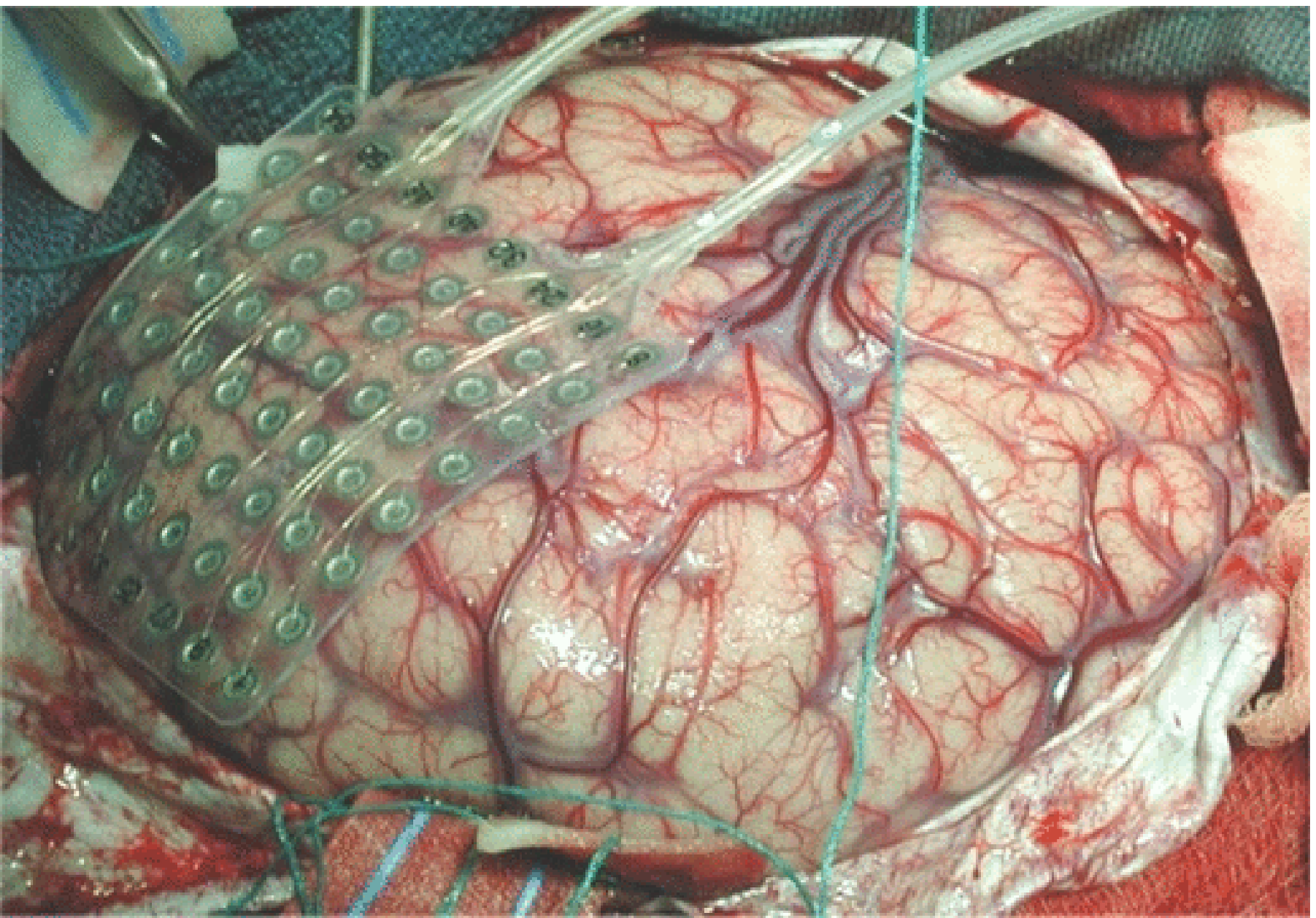,width=4cm,height=3cm,clip=}
\psfig{figure=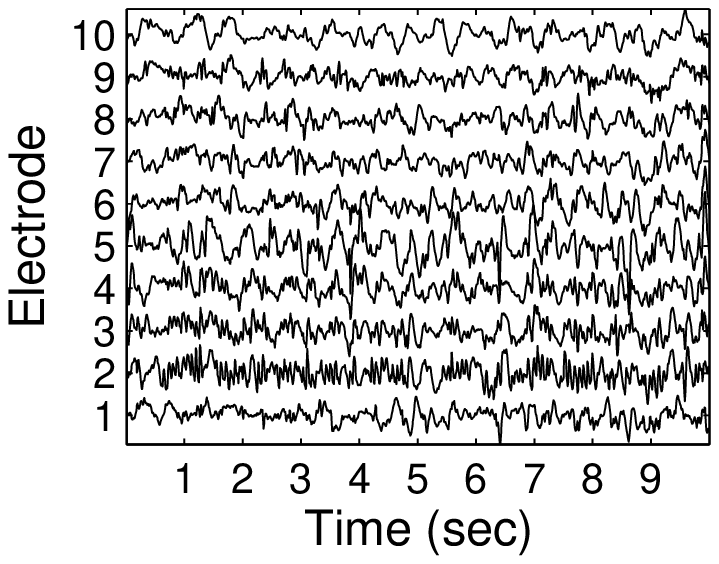,width=4.2cm,height=3cm,clip=}}
\caption[]{\footnotesize : (a) {\bf Subdural EEG (ECoG) grid of
electrodes} placed on the pial surface of the brain for chronic
evaluation of epileptic patients before surgical resection. (b)
{\bf Voltage traces of ECoG signals.} A 10 second time window
display of voltage traces of 10 electrodes, taken from a
multi-electrode recording of 96 electrodes.
 }
\label{figone}
\end{figure}
\begin{figure}[btp]
\centerline{a \hspace{6.2cm} b} \centerline{
\psfig{figure=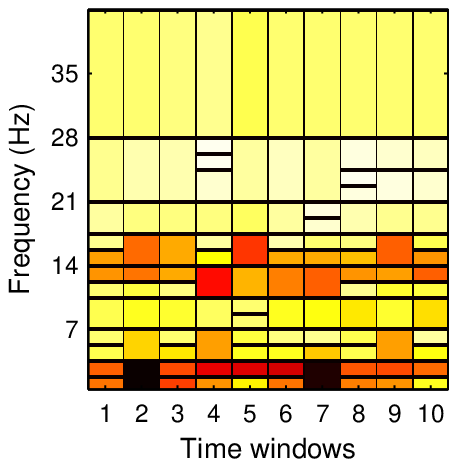,width=4cm,height=4cm,clip=}
\psfig{figure=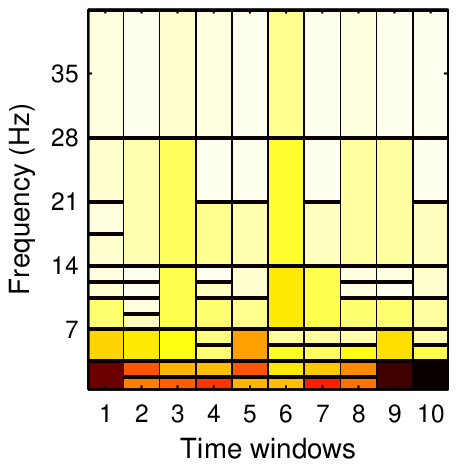,width=4cm,height=4cm,clip=}}
\centerline{c \hspace{6.2cm} d} \centerline{
\psfig{figure=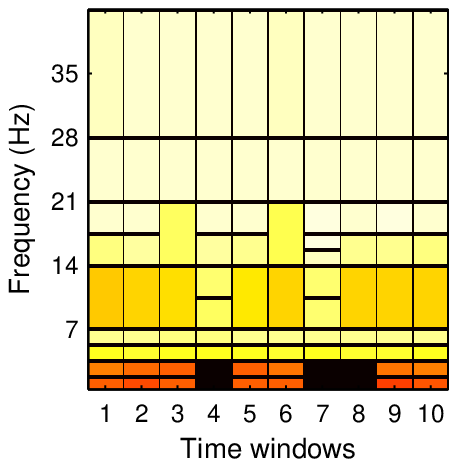,width=4cm,height=4cm,clip=}
\psfig{figure=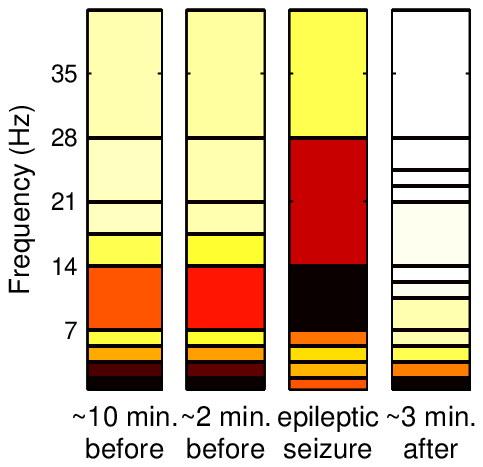,width=4cm,height=4cm,clip=}}
\centerline{e \hspace{6.2cm}} \centerline{
\psfig{figure=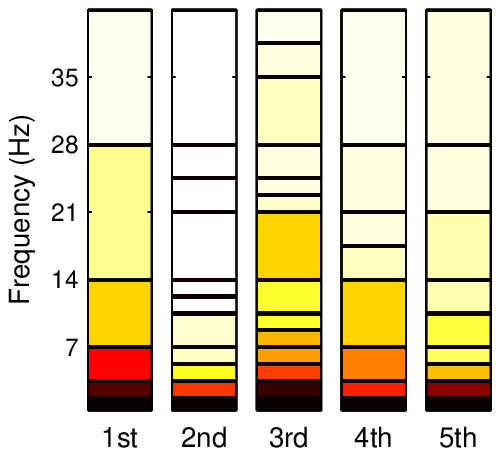,width=6cm,height=4cm,clip=}}
\caption[]{\footnotesize {\bf Personal state-specific frequency
bands.} The frequency subbands division of the best basis of a
single ECoG electrode signal can change over time, as can be seen
in (a) and (b), when calculated at 10 consecutive time windows of
9 seconds each (we used only 5 levels of decomposition, to avoid a
large influence due to a negligent number of coefficients, when
using short time windows). Comparing (a) to (b), it is evident
that the frequency subbands division is also distinct for the
different spatial locations. Even the ensemble best basis of all
the 96 ensemble electrodes, as shown in (c), does not yield
time-invariant partitioning into frequency subbands. However,
temporal coarse-graining of the ensemble best basis by averaging
over time windows of approximately 2.5 minutes produces robust
frequency bands that are time-invariant for long periods(over 10
minutes). This robustness is presented in (d) by examples from the
same recordings of such time-invariant partitioning into frequency
bands (before the seizure) \cite{threehoursapart}. Nonetheless,
the large diversity between individuals, which can be observed in
the 5 examples in (e), leads us to regard them as state-specific
frequency bands partitioning or spectral signatures.
 }
\label{figthree}
\end{figure}

{\bf Introduction.} The various brain tasks (cognitive,
sensory, motor etc.) are performed simultaneously in many
locations and operate at different time scales. In order to
quantify abnormal vs. normal brain function, as in the case of
epilepsy, it is necessary to develop time-invariant templates for
characterization of normal behavior.  The challenge is to decipher
time-invariant features from multi-electrode EEG recordings of
brain activity. Moreover, from Physics perspective, it is not
clear a priori that time-invariant elements do exist since the
activity is inherently
nonergodic.\\
Here we present a new approach of temporal and ensemble averaging
of best-bases constructed from Wavelet Packets Decomposition (WPD)
of the recorded brain activity. The approach is illustrated via
the analysis of subdural EEG (ECoG) recordings from grids of
electrodes that are placed directly on the surface of the cortex
(Fig. \ref{figone}). Such recordings are performed to analyze the
brain activity of epileptic patients under chronic evaluation
before resection surgery to remove the epileptic focus or foci
\cite{epilepsy,Litt2002,Schiff98,Towle2002}. Using this approach,
we discovered the existence of time-invariant personal
state-specific frequency bands above an essential time scale of
about 2.5 minutes. We devised a quantitative measure for
comparison between WPD bases.  Our new person adapted analysis can
help, for example, in the identification of the epileptic foci and
in the development of quantitative analysis methods for early
warning of epileptic seizures. As a self-consistency test that the
frequency bands are not an artifact of the analysis, we show that
the same templates are obtained for subdural and scalp EEG
recordings of the same person.
\\
{\bf The recorded ECoG signals.} The signals analyzed here are
ECoG recordings from grids of typically $\sim$100 electrodes.
These recordings are obtained from epileptic patients undergoing
chronic evaluation for surgery. The electrodes are spatially
distributed over the suspected focal region, so that the focus or
foci could be localized by analyzing the ensemble of signals. The
amplitude of the signals records the electrical voltage at each
electrode (arguably recording local field potentials [LFP]
\cite{Lachaux2003}). The voltage signals are simultaneously
digitized at 112Hz (sampling time interval $\Delta t_{min} \simeq$
9 mSec) with a lowpass filter up to 40Hz. The analysis is usually
performed in time windows of $N_{bin}$ = 1024 samples.\\
{\bf Time-Frequency analysis.} In general, the possible time
intervals for a recorded sequence of $N_{bin}$ elements can range
from $\Delta t_{min} = 1$ (in units of the sampling time interval)
to $\Delta t_{max} = N_{bin}$. In principle, one can extract
information about $N_{bin}$ time intervals at each of the
$N_{bin}$ temporal locations along the sequence. However, such an
$N^{2}_{bin}$ matrix for a sequence of only $N_{bin}$ data samples
must contain redundant information (i.e. over-complete
representation of the recorded sequence). In order to avoid such
redundancy, only $N_{bin}$ time-frequency locations should be
selected, subject to the uncertainty constraint between time and
frequency resolutions - $\Delta t\cdot\Delta f = 1$.\\
Since there are $N_{bin}$ corresponding frequency bands, ranging
from $\Delta f_{min} = 1/N_{bin}$ to $\Delta f_{max} = 1$ (in
units of the Nyquist frequency), each location can be assigned a
local relative resolution $\Delta t / \Delta f$ out of $N_R = 1 +
log_2(N_{bin})$ possible ratios (for simplicity, $N_{bin}$ of the
sequences considered here are in factors of 2). It is convenient
to illustrate both constraints as tiling of the time-frequency
plane with $N_{bin}$ rectangles, each with its own aspect ratio
(height $\Delta f$ and width $\Delta t$), representing the
relative resolutions in time and frequency
\cite{Hulata2004,Hulata2005}.\\
{\bf The WPD as a binary tree.} The Wavelet Packets Decomposition
we use here was devised to partition (tile) the time-frequency
plane into such rectangles (referred to as 'information cells' or
'Heisenberg boxes') \cite{Coifman92,Coifman93,Mallat99}. Each
possible combination of $N_{bin}$ non-overlapping tiles that
geometrically covers the entire corresponding time-frequency plane
can serve as a complete basis spanning the recorded sequence.\\
The WPD is computed by iterating a set of lowpass and highpass
filters (H and G respectively). The functions underlying the
expansions of H and G are "wavelets" ("mother") and "scaling"
("father") functions \cite{Wickerhauser93}. At each iteration, the
wavelet packet coefficients are computed by convoluting the signal
with the filters. Here we utilize the WPD using the 'Coiflet' of
order 1 as a "mother" wavelet (smallest time support of all
'Coiflets')\cite{coiflet,Blanco98,Gutierrez2001,WPDfilterexplanation}.
\begin{figure}[btp]
\centerline{\hspace{0.2cm} a \hspace{3cm} b \hspace{0.7cm} c
\hspace{1.9cm}} \centerline{
\psfig{figure=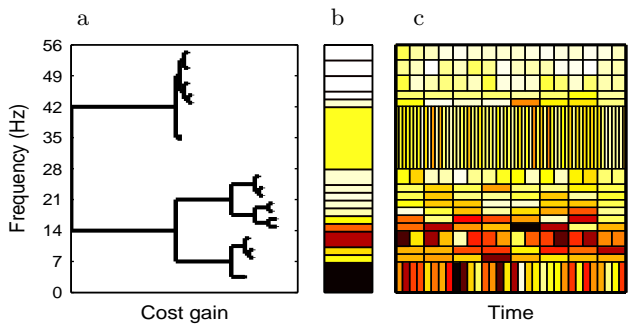,width=8.5cm,height=4cm,clip=}}
\caption[]{\footnotesize {\bf An illustration of the Wavelet
Packets Decomposition (WPD) for a typical signal recorded from a
single electrode.}(a) A binary tree representation of the best
basis obtained by WPD. Each node in the tree represents a basis
frequency band. The vertical axis represents the frequency while
the horizontal axis represents the information cost gain of the
basis frequency band blocks. (b) The corresponding information
distribution in the frequency bands of the best basis (darker
shades represent more information). (c) 2-D tiling representation
of the WPD of the signal (darker shades represent larger wavelet
packet coefficients). The vertical axis represents the frequency
while the horizontal axis represents the time. If the rectangles
were colored according to their information cost contribution then
the information distribution shown in (b) would simply be a sum
over the rows of the basis coefficients.
 }
\label{figtwo}
\end{figure}\\
{\bf Using the Best Basis algorithm.} The WPD generates an
over-complete representation of the signal. The challenge is to
select, out of all possible representations, the one that is the
most efficient in extracting the features of interest. The Best
Basis algorithm is a method for selecting a basis that spans the
signal with a small number of significant packets
\cite{Coifman92,Coifman93}. For that, each wavelet packet function
is assigned an information cost value $M_q(q) = -q\cdot\log_2(q)$
where $q$ is the normalized energy of the wavelet packet. The
total information cost of a frequency subband is obtained by
summing over all the packets in the subband:
\begin{equation}
\label{Msubband}{ M_{subband} = \sum_{k=1}^{K} M_q(q_k) .}
\end{equation}
Viewing the frequency subbands as nodes in a binary tree, the
selection of the best basis is similar to a binary tree search.
Starting from the lowest level bands, we select for each pair of
subbands either the two subbands or their joint "parent" band,
whichever has the lower information cost. The process is repeated
at subsequent levels, going up the scales, back to the global
root. Doing so, we select the set of subbands with the lowest
possible information cost (Fig. \ref{figtwo}).\\
{\bf Bases Similarity Measure.} Every wavelet packets basis can be
described by the frequency subbands partitioning and the
corresponding information cost of every one of the subbands. We
suggest using the following information cost similarity measure,
annotated $ICS$, for comparison between decomposition bases -
$basis1$ and $basis2$:
\begin{equation}
\label{BasSim}{ ICS = \frac{\sum_{n1=n2} (M_{n1} +
M_{n2})}{\sum_{n1}M_{n1} + \sum_{n2}M_{n2}},}
\end{equation}
where $M_{n1}$ and $M_{n2}$ are the information cost of subbands
$n1$ and $n2$ for $basis1$ and $basis2$, respectively. The
summation in the nominator is over all the common subbands of the
two bases. The idea is to compare the information cost included in
the similar subbands to the total information cost. Note, that the
measure assumes values between $0$ (if the bases are totally
dissimilar) and $1$ (if they are exactly the same).\\
{\bf The Ensemble Best Basis of multi-electrodes recordings.} In
Figs. \ref{figthree}a and \ref{figthree}b we show an example of
the evaluated Best Bases for the recordings from two different
electrodes at consecutive time windows ($\sim$ 10 seconds). As can
be seen in these figures, the Best Bases differ from electrode to
electrode ($ICS=0.46\pm0.20$ between these electrodes) and also
vary between consecutive time windows for the same electrode
($ICS=0.58\pm0.12$ [Fig. \ref{figthree}a] and $ICS=0.63\pm0.25$
[Fig. \ref{figthree}b]). Looking for invariant Best Bases, we
proceed to evaluate an ensemble Best Basis for all the electrodes.
Following previously devised methods
\cite{Saito95,Saito96,Saito98,Mallat98}, we average over the
information cost binary tree of each of the $L$ recorded signals
in the ensemble. This is done by first calculating the information
cost of all the nodes in the binary WPD tree for each of the
signals. Next, we evaluate the mean information cost of every node
for the $L$ signals by simple averaging. Then we apply the Best
Basis algorithm to the mean values tree. Thus, we obtain a basis
that may not be optimal for each signal, but rather underlines the
mean content of the ensemble. The nonergodic nature of the brain
activity is reflected in the fact that the resulting ensemble best
basis (EBB) varies between successive short time windows
($ICS=0.65\pm0.33$). Similarly, if we start with temporal
coarse-graining of the Best Bases for the individual electrodes,
the resulting Best Bases are different from each other and from
the Ensemble Best Bases. This is a reflection of the inherent
nonergodicity of the brain activity.
\\
{\bf Time-Invariant Best Bases.} However, we did discover the
existence of an underlying time-invariant ($ICS\simeq1$) Best
Basis in the nonergodic activity. This basis is discovered by
combining ensemble averaging and temporal coarse-graining over a
new essential time scale. Namely, by temporal coarse-graining of
the short time EBBs over a time window wider than an essential
time scale (about 2.5 minutes) \cite{averagingcheck}. The latter
satisfies the requirement that the ICS between EBBs at different
time segments is larger than 0.95.\\
As shown in Fig. \ref{figthree}d, the resulting Best Basis is
time-invariant for recorded periods that are much longer than the
essential time scale \cite{threehoursapart}. Hence, the
time-invariant bases can be used as a frequency decomposition
template for analyzing the recorded brain activity at different
times and locations \cite{Salvador2005}.
\begin{figure}[btp]
\centerline{a \hspace{4cm} b \hspace{2.1cm}} \centerline{
\psfig{figure=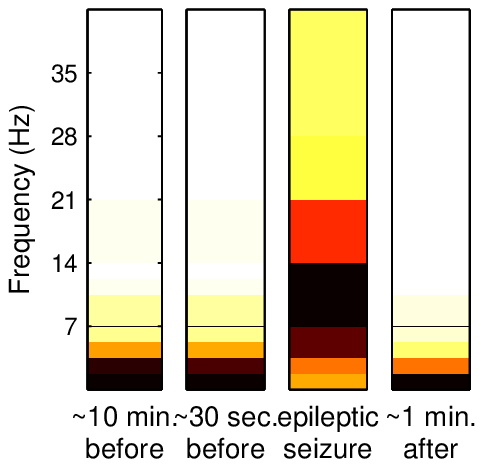,width=4cm,height=4cm,clip=}
\hspace{0.3cm}
\psfig{figure=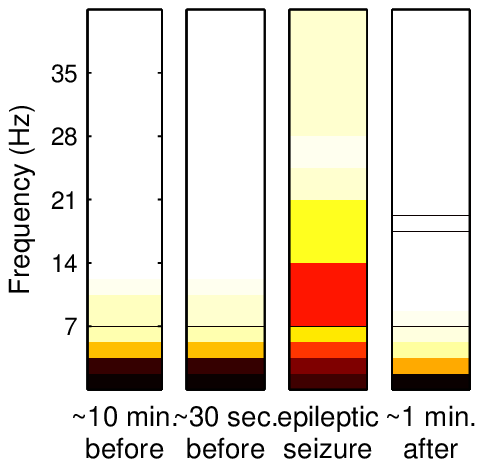,width=4cm,height=4cm,clip=}}
\caption[]{\footnotesize {\bf The frequency templates - before,
during and after epileptic seizure.} (a) The stability of the
frequency template of the ECoG recording is demonstrated for time
intervals of approximately 2.5 minutes each, 10 minutes before the
seizure onset and 30 seconds before the seizure onset. During the
seizure, the energy distribution of the signals changes
dramatically \cite{Krystal99}. After the seizure, the energy
distribution changes again, as the energy in the high frequencies
decreases. (b) Preliminary analysis of scalp EEG signals from the
same recording of (a) shows the same frequency template
(inter-ictal and post-ictal) and the same temporal changes in the
energy distribution.
 }
\label{figfour}
\end{figure}\\
{\bf Personal State-Specific Frequency Bands.} Applying the new
spatio-temporal averaging of the Best Bases to the recordings from
different persons (we analyzed recordings from 12 persons), we
found that each has its own state-specific time-invariant Best
Basis with its own characteristic features, as illustrated in Fig.
\ref{figthree}e. We emphasize that each of these bases bears
resemblance to the classical EEG frequency bands, yet has its own
specific significant deviations from it. Reflecting this notion,
we note that the inter-patient similarity is $ICS=0.75\pm0.16$ for
these examples, which is higher than the similarity between the
variations of an EBB of a single patient over short time windows.
However, this inter-patient similarity is significantly lower than
$ICS\simeq1$, which is measured for bases of different time
windows of recordings of a single patient above an essential time
scale, and could not be considered as invariant. We propose that
the frequency bands are personal specific spectral signatures that
can be used in patient-specific diagnosis of
recorded brain activity.\\
{\bf Self-Consistency Test.} To further substantiate this idea
(and that the calculated frequency bands are not an artifact), we
show a comparison of the frequency bands calculated in parallel,
for the same person, both from ECoG and scalp EEG recordings. As
illustrated in Fig. \ref{figfour}, between seizure episodes
(inter-ictal) the calculated frequency bands are almost identical
($ICS\simeq1$). We also show that the frequency partitioning
changes during the seizure (ictal) episodes. Hence, we expect that
a decomposition of the ictal activity according to the inter-ictal
bands can help in seizure diagnosis.\\
{\bf Conclusions.} These results illustrate the potential value of
the personal best-basis to serve as a template for quantitative
analysis of the epileptic activity \cite{Percha2005}. For example,
the ictal activity can be decomposed according to the inter-ictal
bands during the chronic monitoring of the brain activity and vice
versa. Our new person adapted analysis can help, for example, in
the identification of the epileptic foci. It can also be used to
develop quantitative person-specific analysis for early warning of
epileptic seizures
\cite{Lehnertz98,Arnhold99}.\\
Beyond recorded brain activity, we expect the new approach to be
helpful in revealing the existence of essential time scales and
time-invariant frequency decomposition templates in a wide class
of other nonergodic biological systems with multi-time scale
dynamics. As in the case of the brain, we expect that revealing
such hidden templates can help in analyzing variations in the
systems function and performance.


\begin{thebibliography}{10}
\bibitem{averagingcheck}
\newblock We checked that the way that the averaging is done
(number of time windows, their size etc.) is not important,
provided that it is done over intervals longer than 2.5 minutes.
\end{thebibliography}
\end{document}